\documentclass[reprint,groupedaddress,aps,prl]{revtex4-2}

\usepackage{graphicx}
\usepackage{color}
\usepackage{dcolumn}
\usepackage{bm}
\usepackage{amsmath}
\usepackage{amssymb}
\usepackage{siunitx}
\usepackage{amsfonts}
\usepackage{stmaryrd}
\usepackage{subfigure}
\usepackage{bm}
\usepackage{bbm}
\usepackage[version=4]{mhchem}
\usepackage{tabularx}
\usepackage{multirow}

\usepackage{algpseudocode}
\usepackage{algorithm}

\usepackage{makecell}

\bibpunct{}{}{,}{s}{}{\textsuperscript{,}}  

\newcommand{\CSF}{\mathrm{CSF}}
\newcommand{\DET}{\mathrm{DET}}

\begin{document}

\title{
Entanglement-minimized orbitals enable faster quantum simulation of molecules
}
\author{Zhendong Li}
\email{zhendongli@bnu.edu.cn}
\affiliation{
Key Laboratory of Theoretical and Computational Photochemistry, Ministry of Education, College of Chemistry, Beijing Normal University, Beijing 100875, China
}

\begin{abstract} 
Quantum computation offers significant potential for accelerating the simulation of molecules and materials through algorithms such as quantum phase estimation (QPE). However, the expected speedup in ground-state energy estimation depends critically on the ability to efficiently prepare an initial state with high overlap with the true ground state. For strongly correlated molecules such as iron-sulfur clusters, this overlap is demonstrated
to decay exponentially with system size.
To alleviate this problem, we introduce an efficient
classical algorithm to find entanglement-minimized orbitals (EMOs) using spin-adapted matrix product states (MPS) with small bond dimensions. 
The EMO basis yields a more compact ground-state representation, significantly easing initial state preparation for challenging systems.
Our algorithm improves initial state overlap by nearly an order of magnitude over prior orbital optimization approaches for an iron-sulfur cluster with four irons, and is scalable to larger systems with many unpaired electrons, including the P-cluster and FeMo-cofactor in nitrogenase with eight transition metal centers. For these systems, we achieve substantial enhancements on initial state overlap by factors of $\mathcal{O}(10^2)$ and $\mathcal{O}(10^5)$, respectively, compared to results obtained using localized orbitals. Our results show that initial state preparation for these challenging systems requires far fewer resources than prior estimates suggested.
\end{abstract}

\maketitle



{\it Introduction.---}Strongly correlated systems, such as iron-sulfur clusters in nitrogenase with intricate spin couplings, are notoriously difficult to simulate on classical computers\cite{reiher2017elucidating,li2019electronic2,lee2023evaluating}. 
Even state-of-the-art classical algorithms, such as density matrix renormalization group\cite{white1999ab,chan2011density,szalay2015tensor,baiardi2020density} (DMRG),
struggle to achieve chemical accuracy ($\sim$1 kcal/mol) for these systems. 
Quantum computing offers a promising alternative for simulating electronic structures\cite{cao2019quantum,mcardle2020quantum,bauer2020quantum}.
Algorithms such as quantum phase estimation\cite{kitaev1995quantum,abrams1999quantum,aspuru2005simulated} (QPE) 
provide potential exponential speedups for calculating ground-state energies, 
if a high-quality initial state with substantial overlap with the true ground state can be  prepared efficiently\cite{gharibian2022dequantizing}. 
However, for strongly correlated systems, preparing a good initial state tends to be increasingly difficult as system size grows.
For instance, the weight $p_0\equiv|\langle\Phi|\Psi_0\rangle|^2$ of
either the dominant Slater determinant (DET) $\Phi_{\DET}$ or 
the leading configuration state function (CSF) $\Phi_{\CSF}$,
a spin-adapted linear combination of DETs, in the ground state $\Psi_0$ 
is estimated to decay exponentially from
$\mathcal{O}(10^{-2})$ for iron-sulfur clusters with
two irons to $\mathcal{O}(10^{-7})$ for the FeMo-cofactor (FeMoco) in nitrogenase
containing eight transition metal centers\cite{lee2023evaluating}.
This implies that $\mathcal{O}(10^{7})$ repetitions of the QPE circuit would be required for FeMoco, representing a critical bottleneck for efficient quantum simulations and highlighting the need for improved initial state preparation methods.

Previous efforts to improve initial state preparation have been primarily focusing on preparing more correlated states beyond single DET/CSF, such as linear combinations of DETs\cite{tubman2018postponing,fomichev2024initial}, matrix product states\cite{schon2005sequential,ran2020encoding,malz2024preparation,
fomichev2024initial,berry2025rapid} (MPS), and low-depth parameterized quantum circuits using variational quantum eigensolver (VQE)\cite{peruzzo2014variational,kandala2017hardware}.
However, the choice of orbitals, which is another degree of freedom in quantum simulation of molecules, remains largely unexplored.
Recently, Ollitrault et al.\cite{ollitrault2024enhancing} optimized the orbital basis by maximizing
the weight $|\langle\Phi_{\DET}|\tilde{\Psi}_0\rangle|^2$ of a single determinant with an approximate ground-state wavefunction $\tilde{\Psi}_0$, represented by a truncated set of DETs sampled from MPS. For iron-sulfur clusters with four irons, they achieved an
encouraging enhancement of $p_0$ by two orders of magnitude compared to localized molecular orbitals (LMOs). 
When scaling to larger systems such as FeMoco,
this approach may face two challenges.
First, the truncated approximation to the target state may degrade in quality very quickly as system size increases. Second, the use of $\Phi_{\DET}$ becomes less representative as the number of unpaired electrons increases, since its weight will decay exponentially.
Although using CSFs could mitigate these issues\cite{ollitrault2024enhancing}, the computational cost
will increase prohibitively with the number of unpaired electrons in this framework.
   
In this Letter, we introduce an efficient classical orbital optimization algorithm that significantly enhances the initial state overlap. Rather than directly optimizing the overlap between a DET/CSF and an approximate ground state, our approach minimizes the entanglement entropy
of spin-adapted MPS\cite{mcculloch2002non,sharma2012spin} with small bond dimensions through randomized global orbital optimization, which preserves the SU(2) spin symmetry and is scalable to larger systems with more than four transition metal centers. 
The resulting entanglement-minimized orbitals (EMOs) offer a compact wavefunction representation, benefiting both classical and quantum simulations of strongly correlated systems. For the P-cluster and FeMoco with eight transition metals, we achieve a significant enhancement in initial state overlap by factors of $\mathcal{O}(10^2)$ and $\mathcal{O}(10^5)$, respectively, compared to results obtained using LMOs. Our results indicate that state preparation for these challenging systems is more achievable than previously believed, paving
the way for advancing both classical and quantum simulation techniques for strongly correlated systems.

\begin{figure}
\includegraphics[width=0.48\textwidth]{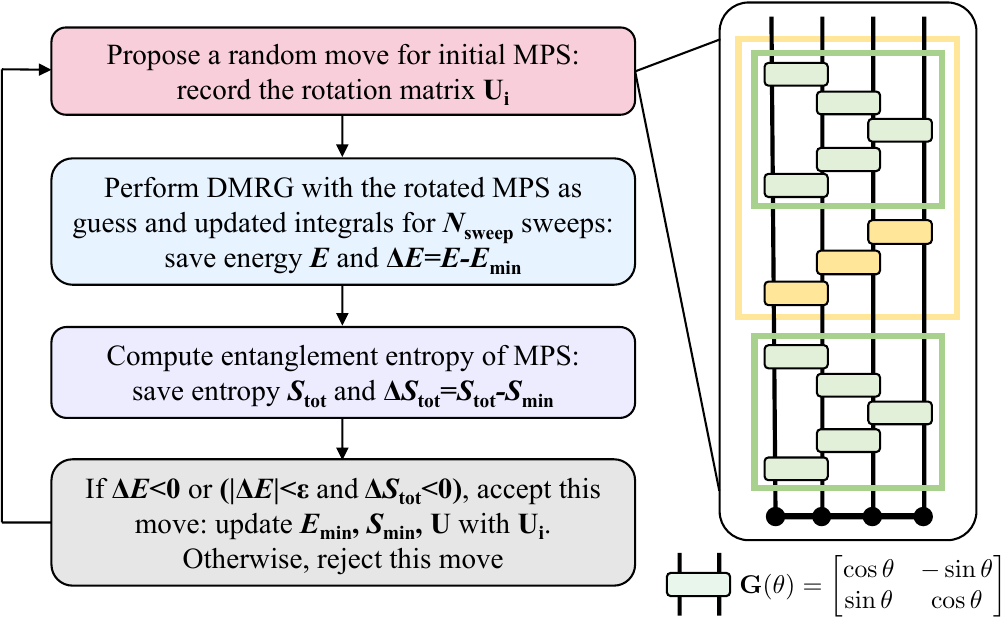}
\caption{Flowchart of the randomized global orbital optimization algorithm for finding EMOs by
simultaneously minimizing energy $E$ and total entropy $S_{\mathrm{tot}}$. 
The MPS is represented by a chain of black dots, each corresponding to a site tensor $A[k]$. The core of the algorithm involves proposing a random orbital rotation, 
which consists of locally optimized Givens rotations (green gates) aimed at
minimizing the R\'{e}nyi entropy \eqref{eq:renyi} with $\alpha=1/2$, as well
as random swap layers (yellow gates). The later is implemented by selecting $\theta$ in the Givens rotations from the set $\{0,\pi/2\}$ with an equal probability. The green boxes indicate the gates are repeated until $S_{\mathrm{tot}}$ is minimized, while the yellow box represents the repeated execution of the random swap layer and local optimization for a predefined number of steps $N_{\mathrm{macro}}$ to ensure the generation of a random move with an appropriate step size.
}\label{fig:scheme}
\end{figure}

{\it Algorithm for entanglement-minimized orbitals.---}We leverage the MPS representation for approximating the ground state of molecules, which has been very successful for strongly correlated systems\cite{white1999ab,chan2011density,szalay2015tensor,baiardi2020density}. 
Specifically, for systems with $K$ spatial orbitals (sites), MPS can be written as
\begin{eqnarray}
|\Psi\rangle = \sum_{n,\alpha} A^{n_1}_{\alpha_1}[1]
A^{n_2}_{\alpha_1\alpha_2}[2]
\cdots A^{n_K}_{\alpha_{K-1}}[K]|n_1n_2\cdots n_K\rangle,\label{eq:MPS}
\end{eqnarray}
where $A^{n_k}_{\alpha_{k-1}\alpha_k}[k]$ are site tensors and
$n_k\in\{|0\rangle,|\alpha\rangle,|\beta\rangle,|2\rangle\}$.
It has a simple diagrammatic representation\cite{orus2014practical} shown in Fig. \ref{fig:scheme}. 
As mentioned before, instead of directly optimizing the overlap between a DET/CSF and MPS, 
we design an orbital optimization algorithm to minimize the entanglement entropy along the MPS chain for two reasons. 
First, rotating an MPS to a new orbital basis or computing the overlap
between an rotated DET/CSF with MPS is computationally demanding.
In fact, calculating the overlap between an arbitrary DET with an MPS up to a multiplicative error is proven to be \#P-hard\cite{jiang2025unbiasing}.
Second, we observe that by minimizing the entanglement, a more compact representation of 
wavefunction can be obtained, which usually leads to a large weight for the dominant DET/CSF. Although this may not be optimal for achieving the maximal $p_0$, 
it can already be satisfactory for the purpose of initial state preparation for QPE. Besides, in the limiting case, where the ground state is a DET, minimizing the entanglement of the ground state and
maximizing the overlap between a DET and the ground state will yield the same DET. The advantage of minimizing entanglement is that it can be efficiently carried out using sweep optimization\cite{schollwock2011density} for MPS.

Consider the minimization of the entanglement between the first orbital and the rest of the orbitals as an example. We begin by expressing the MPS \eqref{eq:MPS} as
\begin{eqnarray}
\Psi^{n_1n_2\cdots n_K} = \sum_{\alpha_2} C^{n_1n_2}_{\alpha_2} R^{n_3\cdots n_K}_{\alpha_2},\label{eq:bipartition12}
\end{eqnarray}
where $R^{n_3\cdots n_K}_{\alpha_2}$ satisfies the so-called right canonical condition\cite{schollwock2011density} $\sum_{n} R^{n_3\cdots n_K *}_{\alpha_2}R^{n_3\cdots n_K}_{\alpha_2'}=\delta_{\alpha_2\alpha_2'}$.
Then, the Schmidt decomposition is performed by the singular value decomposition (SVD) of
the matrix $M_{n_1,n_2\alpha_2}\equiv C^{n_1n_2}_{\alpha_2}$. The Schmidt decomposition at other bipartition positions can be performed by transforming Eq. \eqref{eq:bipartition12}
successively using SVD\cite{schollwock2011density}. In general, we denote the obtained singular values for the bipartition at the position $(k,k+1)$ by $\{\lambda_i[k]\}$ satisfying $\sum_i\lambda_i^2[k]=1$. The corresponding R\'{e}nyi entropy of order $\alpha$ is 
\begin{eqnarray}
S_{\alpha}[k] = \frac{1}{1-\alpha}\log\left(\sum_i \lambda_i^{2\alpha}[k]\right).\label{eq:renyi}
\end{eqnarray}
Our goal is to design an algorithm to minimize the sum of R\'{e}nyi entropies along the MPS chain, i.e., $S_{\mathrm{tot}} = \sum_{k=1}^{K-1}S_{\alpha}[k]$, by rotating orbitals, while approximating the ground state by minimizing the energy of MPS. 

In the context of mode optimization for DMRG, Krumnow
et al.\cite{krumnow2016fermionic} applied local orbital rotations 
to minimize $S_{\mathrm{tot}}$ in a sweep fashion, 
see the lower green box in Fig. \ref{fig:scheme}. 
Specifically, at the bipartition position \eqref{eq:bipartition12}, a Givens rotation is applied to the first two orbitals
\begin{eqnarray}
(\tilde{\hat{a}}_{1\sigma}^\dagger,\tilde{\hat{a}}_{2\sigma}^\dagger)
=
(\hat{a}_{1\sigma}^\dagger,\hat{a}_{2\sigma}^\dagger)
\mathbf{G}(\theta),\;\;
\mathbf{G}(\theta)=
\begin{bmatrix}
\cos\theta & -\sin\theta \\
\sin\theta & \cos\theta
\end{bmatrix},\label{eq:givens}
\end{eqnarray}
where $\hat{a}_{p\sigma}^{(\dagger)}$ are the Fermionic annihilation (creation) operator. This will induce a change on the wavefunction amplitude (see Supplementary Material\cite{SM} for details), which is shown pictorially
in Fig. \ref{fig:scheme} by the application of a rotation gate (green block) on top of the MPS. The parameter $\theta$ is determined by minimizing $S_{\alpha=1/2}$ for the rotated wavefunction amplitude. Unfortunately, this local optimization scheme can easily get stuck in local minima\cite{friesecke2024global}, since only nearest neighbor orbitals are allowed to mix. To overcome this problem, we develop a randomized global orbital optimization algorithm by combining local optimization with random perturbation, based on the idea of the basin hopping algorithm\cite{wales1997global}, a stochastic global optimization algorithm widely used to find the lowest-energy structures for clusters in chemistry and physics.
The flowchart of our algorithm is shown in Fig. \ref{fig:scheme}, and the implementation details are given in Supplementary Material\cite{SM}. The outline of our algorithm for finding EMOs is as follows.

Prior to orbital optimization, we assume that an initial MPS, with a small bond dimension $D$ for truncating $\{\alpha_k\}$, has been obtained using 
spin-adapted DMRG\cite{mcculloch2002non,sharma2012spin} for the electronic Hamiltonian $\hat{H}$ in the initial orbital basis
\begin{eqnarray}
\hat{H}
=
\sum_{pq,\sigma}h_{pq}\hat{a}_{p\sigma}^\dagger \hat{a}_{q\sigma}
+
\frac{1}{2}\sum_{pqrs,\sigma\tau}
\langle pq|rs\rangle
\hat{a}_{p\sigma}^\dagger \hat{a}_{q\tau}^\dagger
\hat{a}_{s\tau} \hat{a}_{r\sigma},\label{eq:Hamiltonian}
\end{eqnarray}
where $\sigma,\tau\in\{\alpha,\beta\}$ represents spin indices
and $h_{pq}$ ($\langle pq|rs\rangle$) are the one-electron
(two-electron) molecular integrals\cite{helgaker2014molecular}.
Given the initial MPS, we introduce a random orbital rotation, which is designed to involve two basic components, local minimization and random perturbation, applied in an alternating fashion:

(1) We first perform the aforementioned local minimization of entanglement using Givens rotations\cite{krumnow2016fermionic} until $S_{\mathrm{tot}}$ converges. 

(2) We then apply a random perturbation to orbitals and MPS, for which we choose to randomly apply 
swap gates in a linear layout (see Fig. \ref{fig:scheme}). This is simply realized by applying the Givens rotation with $\theta$ chosen from $\{0,\pi/2\}$ with an equal probability. 

(3) After the random perturbation, we apply the local minimization again to prevent
the random swaps creating too large perturbation, leading to
low acceptance in the later stage of the algorithm.

Steps (2) and (3) are repeated for $N_{\mathrm{macro}}$ times ($N_{\mathrm{macro}}=5$ in this work) to complete the entire random move. 
This repetition ensures that sufficiently large steps are taken to escape local minima. We save the orbital rotation matrix, denoted by $\mathbf{U}_i$,  
and the rotated MPS as the initial guess for the DMRG calculations in the new orbital basis. A cutoff $\chi=2D$ for the bond dimension is used in compressing the application of Givens rotation gates to the MPS.

With the rotated orbitals, we perform DMRG calculations
for a small number of sweeps ($N_{\mathrm{sweep}}=4$ in this work) to variationally minimize the energy $E=\langle\Psi|\hat{H}|\Psi\rangle$ in the new basis. After obtaining the optimized MPS, we compute its corresponding $S_{\mathrm{tot}}$. The energy $E$ and entropy $S_{\mathrm{tot}}$ are used to decide whether the current random orbital rotation is accepted or rejected. Specifically, if $E$ is lowered or $E$ is slightly increased (within a tolerance controlled by a parameter $\varepsilon$) but $S_{\mathrm{tot}}$ is reduced, the current orbital rotation is accepted. 

Following the basin hopping algorithm\cite{wales1997global}, the entire procedure, comprising random perturbation, local optimization, and accept/reject decision, is repeated for $N_{\max}$ iterations to produce the final MPS and the associated orbitals with coefficients $\mathbf{U}$. 
The computational scaling for finding EMOs is 
$\mathcal{O}(N_{\mathrm{max}}(K^3D^3+K^4D^2))$.
Finally, to find the largest weight $p_0$ for DET/CSF in an MPS, we perform perfect sampling\cite{white2009minimally,stoudenmire2010minimally,guo2018communication} 
or exhaustive search with pruning\cite{lee2021externally}. 
The present algorithm allows us to explore the energy and entanglement entropy landscapes efficiently, as the random swap layers introduce random permutations of orbitals, which help to escape from local minima. 
Compared to previous work\cite{ollitrault2024enhancing}, 
where an MPS must first be approximated by a sum of DETs,
our algorithm operates entirely within the MPS framework, 
free from such truncation and fully preserving SU(2) spin symmetry\cite{mcculloch2002non,sharma2012spin} (see Supplementary Material\cite{SM} for details). 
This makes it particularly well-suited for strongly correlated systems with many unpaired electrons, where truncated expansions may fail to faithfully represent the MPS due to the exponential increase in the number of important determinants.

{\it Results.---}We employ the iron-sulfur clusters, including [2Fe-2S], [4Fe-4S], P-cluster, and FeMoco, examined in previous works\cite{lee2023evaluating,ollitrault2024enhancing}
as the most challenging systems, to demonstrate the performance of our algorithm.
The active spaces are (30e,20o), (54e,36o), (114e,73o), and (113e,76o) for [2Fe-2S]\cite{li2017spin}, [4Fe-4S]\cite{li2017spin}, P-cluster\cite{li2019electronic}, and FeMoco\cite{li2019electronic2}, respectively, where ($m$e,$n$o) denotes $m$ electrons distributed in $n$ active orbitals. Geometric structures, compositions of active spaces, and other computational details are given in Supplementary Materials\cite{SM}.

\begin{figure}[H]
\centering
\includegraphics[width=0.49\textwidth]{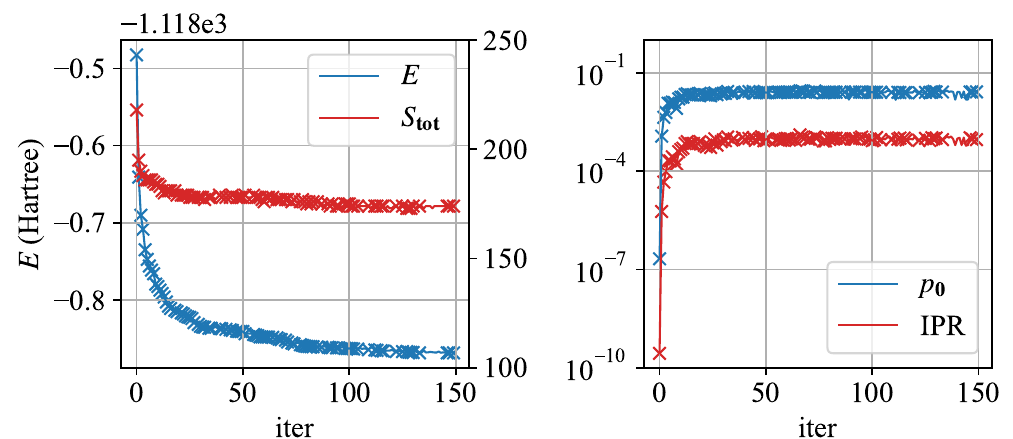}
\caption{
Randomized orbital optimization process for 
the (113e,76o) active space model of FeMoco 
using $D=100$ and $N_{\mathrm{max}}=150$.
Left: $E$ (in Hartree) and $S_{\mathrm{tot}}$. Right: the largest weight of CSF ($p_0^\CSF$) and the inverse participation ratio (IPR). The cross markers indicate the accepted steps.
}\label{fig:femoco}
\end{figure}

We first apply the randomized orbital optimization algorithm with $D=100$ to obtain EMOs. 
Figure \ref{fig:femoco} illustrates the optimization process for FeMoco. Notably, $p_0^\CSF$ 
increases rapidly from $\mathcal{O}(10^{-7})$ to a plateau of $\mathcal{O}(10^{-2})$ within the first 10 iterations of the orbital optimization, which is faster than the convergence for $E$ and $S_{\mathrm{tot}}$. 
The faster convergence of $p_0$ is also observed
for other systems (see Supplementary Material\cite{SM}),
and is consistent with the findings in previous studies\cite{tubman2018postponing,ollitrault2024enhancing}. Furthermore, to measure 
the overall spread of the wavefunction, we compute
the inverse participation ratio (IPR) of the obtained MPS, viz., $\mathrm{IPR}=\sum_{n}|\Psi^{n_1n_2\cdots n_K}|^4$, by Monte Carlo sampling.
As shown in Fig. \ref{fig:femoco}, during the orbital optimization, the wavefunctions exhibit an transition from a delocalized state, with an IPR of $\mathcal{O}(10^{-10})$, 
to a much more compact state with an IPR of $\mathcal{O}(10^{-3})$. 

\begin{figure}
\centering
\includegraphics[width=0.46\textwidth]{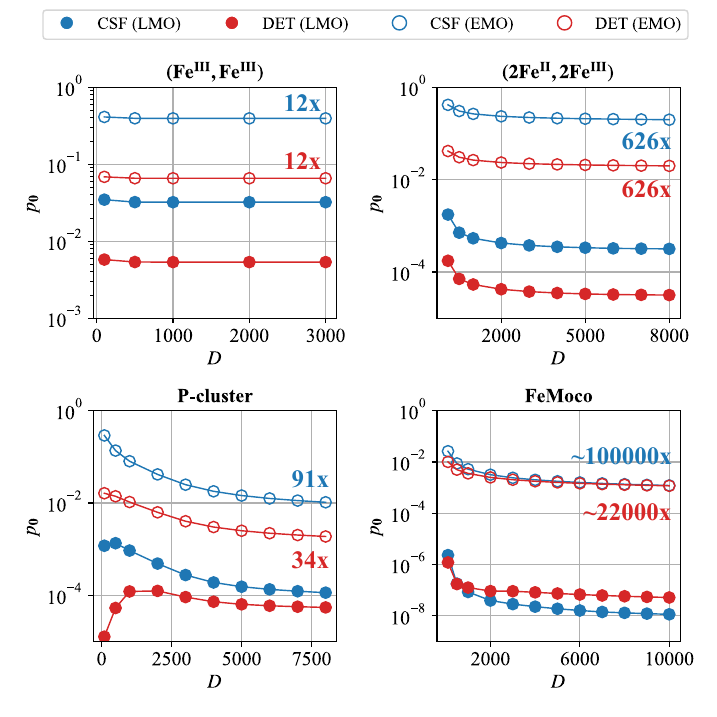}
\caption{
The largest weight $p_0$ of CSF/DET as a function of $D$ obtained by DMRG calculations for iron-sulfur clusters ([2Fe-2S], [4Fe-4S], P-cluster, and FeMoco)
using LMOs (solid circle) and EMOs (open circle).
The enhancement factor on $p_0$ at the largest $D$ is shown for each case.
}\label{fig:p0plot}
\end{figure}

With the optimized EMOs, we then perform large-scale DMRG calculations with increasing $D$ to achieve a more accurate approximation of the ground state. 
The recently developed GPU-accelerated ab initio DMRG implementation\cite{xiang2024distributed}
enables us to push the calculations up to $D=10000$ for FeMoco, establishing a new benchmark for this system.
Figure \ref{fig:p0plot} displays the obtained $p_0$ for CSF/DET as a function of $D$ using LMOs and EMOs. 
It is demonstrated that the relative compactness of the wavefunction in the EMO basis is largely maintained as $D$ increases, since the decay of $p_0$ is no faster than the decay in the LMO basis. Consequently, significant improvements on $p_0$ over the LMO basis are still achieved at large $D$. In particular, we observe a remarkable increase in $p_0^\CSF$ by a factor of approximately $10^5$ at $D=10000$ for FeMoco. Besides, the DMRG energy in the EMO basis is found to be significantly lower than that in the LMO basis for the same $D$ (see Table \ref{tab:femoco}\cite{SM}), demonstrating that the EMO basis is beneficial not only for quantum computing but also
for classical calculations.
As detailed in the Supplementary Material\cite{SM},
the advantages of EMOs extend to other strongly correlated systems such as the two-dimensional Hubbard model.

\begin{figure}
\centering
\includegraphics[width=0.48\textwidth]{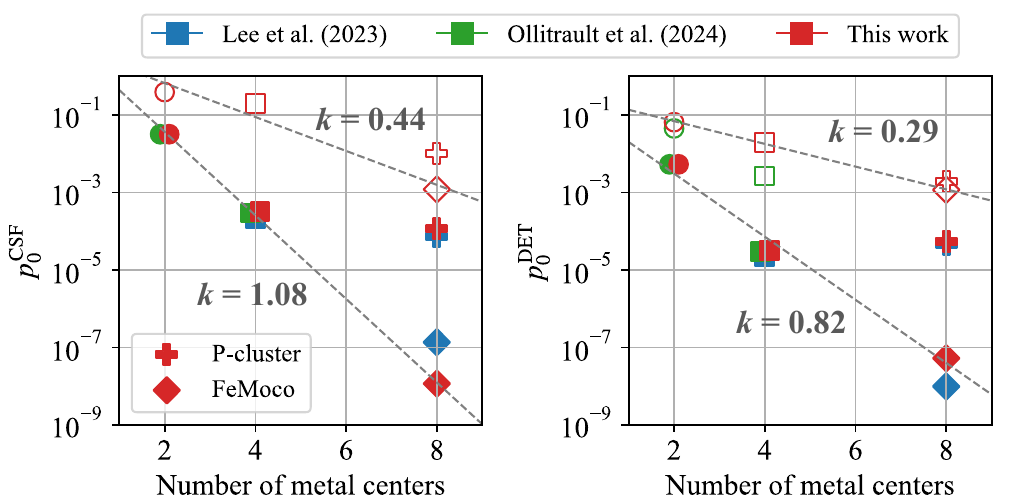}
\caption{Comparison of $p_0$ obtained at the largest $D$
as a function of the number of metal centers $N$
in iron-sulfur clusters with previous studies\cite{lee2023evaluating,ollitrault2024enhancing}.
The fitted lines $p_0=Ae^{-kN}$ (dashed) and exponents $k$ using the data for [2Fe-2S], [4Fe-4S], and FeMoco are shown for a better comparison of the LMO (solid markers) and EMO (empty markers) results. The positions of the 
LMO results for [2Fe-2S] and [4Fe-4S] are slightly
shifted horizontally for better visibility.
}\label{fig:fescluster}
\end{figure}

Finally, we compare our results with previous studies\cite{lee2023evaluating,ollitrault2024enhancing} by plotting $p_0$ as a function of the number of metal centers $N$ in Fig. \ref{fig:fescluster}.
Except for FeMoco, where the difference in $p_0$ is slightly larger,
our results for $p_0$ in the LMO basis align well with previous
results\cite{lee2023evaluating,ollitrault2024enhancing}.
For FeMoco, the discrepancy is likely due to convergence to different low-energy competing
states. Specifically, we find that the previously reported DMRG energy\cite{lee2023evaluating} obtained at the largest bond dimension $D=7000$ is higher by about 28 milli-Hartree
than the present DMRG energy at the same $D$ (see Supplementary Material\cite{SM}).
For the [2Fe-2S] cluster, $p_0^{\DET}$ in the EMO basis is 1.4 times larger than
the optimized result by Ollitrault et al.\cite{ollitrault2024enhancing}, while for the
[4Fe-4S] cluster, $p_0^\DET$ in the EMO basis improves the previous optimized
result\cite{ollitrault2024enhancing} by a factor of 7.5, highlighting the advantage of our algorithm even for small clusters.
As discussed by Lee et al.\cite{lee2023evaluating}, $p_0$ decays exponentially with respect to $N$, presenting a severe obstacle to realizing practical quantum advantage in simulating strongly correlated systems.
Although using EMOs alone cannot completely eliminate this
exponential decay, it substantially mitigates the decay rate, as evidenced in Fig. \ref{fig:fescluster}.
We expect that further improvements can be achieved by integrating EMOs with advanced state preparation techniques
using correlated wavefunctions\cite{tubman2018postponing,fomichev2024initial,schon2005sequential,ran2020encoding,malz2024preparation,
fomichev2024initial,berry2025rapid,peruzzo2014variational,kandala2017hardware}.

{\it Conclusion.---}We introduce the entanglement-minimized orbitals to alleviate the exponential decay of initial state overlap for strongly correlated systems, a key challenge for achieving practical quantum advantage in simulating complex systems. We demonstrate that EMOs can be efficiently obtained
using the randomized orbital optimization algorithm with 
low bond dimension spin-adapted MPS. 
The wavefunction in the resulting EMO basis is highly compact, making it easier to prepare on quantum computers. Our algorithm outperforms previous orbital optimization methods, even for
small iron-sulfur clusters, and provides significant enhancements on $p_0$
for the P-cluster and FeMoco by two and five orders of magnitude, respectively.
By enabling more efficient initial state preparation, 
our approach facilitates the practical application of quantum algorithms such as QPE and VQE, particularly for simulating large, strongly correlated systems.
Moreover, this advancement also benefits classical simulation methods such as DMRG 
for achieving more accurate ground state energies. The initial step in this direction is very promising by combining EMOs with Clifford-augmented MPS\cite{qian2024augmenting,jiale2025augmenting}. 

\section{Acknowledgments}
The author acknowledges helpful discussion with Mingpu Qin and Jiajun Ren. 
This work was supported by the Innovation Program for Quantum Science and Technology (Grant No. 2023ZD0300200) and the Fundamental Research Funds for the Central Universities.

\bibliographystyle{apsrev4-1}
\bibliography{main}

\pagebreak
\clearpage
\raggedbottom
\pagebreak
\widetext
\begin{center}
\textbf{\large Supplemental Material for: \\
Entanglement-minimized orbitals enable faster quantum simulation of molecules
}\\
\vspace{2ex}
Zhendong Li \\
\vspace{2ex}
{\it
Key Laboratory of Theoretical and Computational Photochemistry, Ministry of Education, College of Chemistry, Beijing Normal University, Beijing 100875, China
}
\end{center}

\setcounter{secnumdepth}{3}
\setcounter{section}{0}
\setcounter{equation}{0}
\setcounter{figure}{0}
\setcounter{table}{0}
\setcounter{page}{1}
\makeatletter
\renewcommand{\theequation}{S\arabic{equation}}
\renewcommand{\thesection}{S\arabic{section}}
\renewcommand{\thefigure}{S\arabic{figure}}
\renewcommand{\thetable}{S\arabic{table}}


\section{Implementation details for the randomized global orbital optimization algorithm}
The pseudo code for the randomized global orbital optimization algorithm is presented
in Algorithm \ref{oodmrg}. Here, we provide more details for the implementation.

\begin{algorithm}[H]
\caption{Randomized global orbital optimization in conjunction with DMRG for finding EMOs}
\begin{algorithmic}[1]
\Require an initial MPS, cutoff for bond dimension $D$, threshold $\varepsilon=10^{-8}$, number of repeated random and disentangling layers $N_{\mathrm{macro}}=5$, maximal number of iterations $N_{\mathrm{max}}$,
number of DMRG sweeps $N_{\mathrm{sweep}}=4$, initial orbital rotation matrix $\mathbf{U}=\mathbf{I}$
\Ensure energy-minimized MPS, optimized orbitals specified by $\mathbf{U}$

\State initialize the minimal energy $E_{\min}$ and the sum of entropies $S_{\min}$ to a large positive number.
\For{$i=1$ to $N_{\mathrm{max}}$}

    \State \textbf{propose a random move}: apply initial disentangling layers and $N_{\mathrm{macro}}$ subsequent random and disentangling layers to the MPS, record the corresponding orbital rotation matrix $\mathbf{U}_i$

    \State \textbf{integral transformation}: transform molecular integrals to the rotated basis
    
    \State \textbf{MPS optimization}: perform two-dot DMRG calculations with the new integrals and the rotated MPS as initial guess for $N_{\mathrm{sweep}}$ sweeps, save the final two-dot DMRG energy $E$ and compute the change of energy $\Delta E = E - E_{\min}$

    \State \textbf{measure entanglement}: compute the entropies of the obtained MPS, save $S_{\mathrm{tot}}=\sum_{k=1}^{K-1}S_{1/2}[k]$ and compute $\Delta S_{\mathrm{tot}} = S_{\mathrm{tot}} - S_{\min}$
    
    \State \textbf{accept or reject the move}:
    \If{$\Delta E< 0$ or ($|\Delta E| < \varepsilon$ and $\Delta S_{\mathrm{tot}} < 0$)}
        \State accept: $E_{\min}=E$, $S_{\min}=S_{\mathrm{tot}}$, save MPS for the next iteration, update $\mathbf{U}$ with $\mathbf{U}_i$.
    \Else
        \State reject: the initial MPS will still be used for the next iteration
    \EndIf
\EndFor
\end{algorithmic}
\label{oodmrg}
\end{algorithm}

\subsection{Induced rotation for wavefunction amplitude} 
The basic function in implementing the randomized global orbital optimization algorithm is to apply
a Givens rotation gate on MPS. Suppose the two orbitals are rotated as in Eq. \eqref{eq:givens}. The wavefunction in the old and new orbital bases can be expressed as
\begin{eqnarray}
|\Psi\rangle = \sum_{n} |n_1n_2\cdots n_K\rangle \Psi^{n_1n_2\cdots n_K}
= \sum_{m} |\tilde{m}_1\tilde{m}_2m_3\cdots m_K\rangle \tilde{\Psi}^{m_1m_2\cdots m_K},
\end{eqnarray}
where we used tilde to represent the quantity in the new basis. The wavefunction amplitude in the new basis can be found as
\begin{eqnarray}
\tilde{\Psi}^{\tilde{m}_1\tilde{m}_2n_3\cdots n_K}
=
\sum_{n_1n_2}
\langle\tilde{m}_1\tilde{m}_2|n_1n_2\rangle
\Psi^{n_1n_2n_3\cdots n_K}.
\end{eqnarray}
Let $|\tilde{m}_1\tilde{m}_2\rangle=\sum_{n_1n_2}|n_1n_2\rangle \mathcal{G}_{n_1n_2,m_1m_2}(\theta)$ with $\mathcal{G}$ is an orthogonal matrix that can be computed from
the definition \eqref{eq:givens}, then $\langle\tilde{m}_1\tilde{m}_2|n_1n_2\rangle=
  \mathcal{G}_{n_1n_2,m_1m_2}^*(\theta)=
 [\mathcal{G}^\dagger(\theta)]_{m_1m_2,n_1n_2}=\mathcal{G}_{m_1m_2,n_1n_2}(-\theta)$
 such that
\begin{eqnarray}
\tilde{\Psi}^{\tilde{m}_1\tilde{m}_2n_3\cdots n_K}
=
\sum_{n_1n_2}
\mathcal{G}_{m_1m_2,n_1n_2}(-\theta)
\Psi^{n_1n_2n_3\cdots n_K}.\label{eq:WavefunctionRotation}
\end{eqnarray}
The presence of $-\theta$ maintains $|\Psi\rangle$ unchanged.

\begin{figure}[H]
\centering
\includegraphics[width=0.55\textwidth]{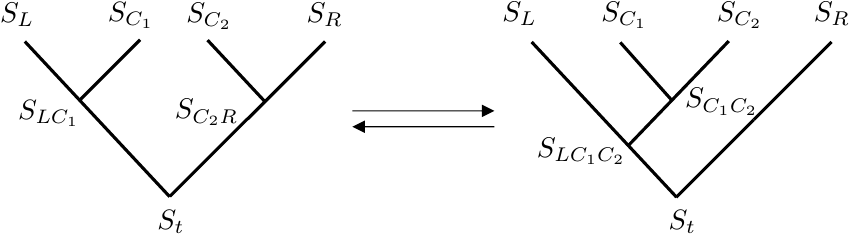}
\caption{
Recoupling of the four angular momenta ($S_L$, $S_{C_1}$, $S_{C_2}$, and $S_R$) 
for the two-dot wavefunction in spin-adapted DMRG\cite{mcculloch2002non,sharma2012spin} required for computing the rotated MPS with the SU(2) spin symmetry.
}\label{fig:recoupling}
\end{figure}

Now we derive the matrix elements of $\mathcal{G}_{m_1m_2,n_1n_2}(\theta)$.
Let $\mathcal{H}_1=\{|0\rangle,|\alpha\rangle,|\beta\rangle,|2\rangle\}$, the
direct product space $\mathcal{H}_2=\mathcal{H}_1\otimes\mathcal{H}_1$ can
be decomposed into a direct sum of different subspaces $\mathcal{H}^{(N_\alpha,N_\beta)}$
with different numbers of $\alpha$ and $\beta$ electrons, viz.,
\begin{eqnarray}
\mathcal{H}_2 &=& \mathcal{H}^{(0,0)}_{\dim=1}\oplus \mathcal{H}^{(2,0)}_{\dim=1} \oplus  \mathcal{H}^{(0,2)}_{\dim=1} \oplus  \mathcal{H}^{(2,2)}_{\dim=1} \nonumber\\
& \oplus & \mathcal{H}^{(1,0)}_{\dim=2}\oplus\mathcal{H}^{(2,1)}_{\dim=2}\oplus
\mathcal{H}^{(0,1)}_{\dim=2}\oplus\mathcal{H}^{(1,2)} _{\dim=2}\nonumber\\
& \oplus & \mathcal{H}^{(1,1)}_{\dim=4}.\label{eq:H1H1}
\end{eqnarray}
There can be three cases:

\begin{description}
\item[Case 1] For the four one-dimensional subspaces in the first line of Eq. \eqref{eq:H1H1}, the basis function is invariant under the orbital rotation. For instance, $\tilde{\mathcal{H}}^{(2,0)}=\{|\tilde{\alpha}\tilde{\alpha}\rangle=|\alpha\alpha\rangle\}$, which follows from
\begin{eqnarray}
|\tilde{\alpha}\tilde{\alpha}\rangle &=& 
\tilde{a}_{1\alpha}^\dagger
\tilde{a}_{2\alpha}^\dagger|vac\rangle
=
(a_{1\alpha}^\dagger c + a_{2\alpha}^\dagger s)
(-a_{1\alpha}^\dagger s + a_{2\alpha}^\dagger c)|vac\rangle 
=
a_{1\alpha}^\dagger a_{2\alpha}^\dagger |vac\rangle
=
|\alpha\alpha\rangle,
\end{eqnarray}
where we used a shorthand notation $c=\cos\theta$ and $s=\sin\theta$ for brevity.

\item[Case 2] The four subspaces in the second line of Eq. \eqref{eq:H1H1} are all two-dimensional. Their basis functions follow the similar transformation rules, viz.,
\begin{eqnarray}
(|\tilde{\sigma}0\rangle,|0\tilde{\sigma}\rangle) 
&=& 
(|\sigma 0\rangle,|0 \sigma\rangle) \mathbf{G}(\theta), \\
(|\tilde{\sigma}\tilde{2}\rangle,|\tilde{2}\tilde{\sigma}\rangle) 
&=& 
(|\sigma 2\rangle,|2\sigma\rangle) \mathbf{G}(\theta).
\end{eqnarray}

\item[Case 3] $\mathcal{H}^{(1,1)}$ is four dimensional and the basis functions
transform as
\begin{eqnarray}
(|\tilde{2}0\rangle,|0\tilde{2}\rangle,|\tilde{\alpha}\tilde{\beta}\rangle,|\tilde{\beta}\tilde{\alpha}\rangle)
=
(|20\rangle,|02\rangle,|\alpha\beta\rangle,|\beta\alpha\rangle)
\begin{bmatrix}
c^2  &  s^2 & -cs & cs  \\
s^2  &  c^2 & cs  & -cs \\
cs   &  -cs & c^2 & s^2 \\
-cs  &  cs  & s^2 & c^2 \\
\end{bmatrix}.\label{eq:transform}
\end{eqnarray}
\end{description}
The above equations give the nonzero elements of $\mathcal{G}_{m_1m_2,n_1n_2}(\theta)$,
which suffices to compute the rotated wavefunction amplitudes using
Eq. \eqref{eq:WavefunctionRotation} or the rotated MPS using U(1) symmetries.
For rotating MPS with the SU(2) spin symmetry, we further need to transform 
the basis functions for $\mathcal{H}^{(1,1)}$ into spin eigenfunctions, viz.,
\begin{eqnarray}
(|20\rangle,|02\rangle,\frac{1}{\sqrt{2}}(|\alpha\beta\rangle-|\beta\alpha\rangle),
\frac{1}{\sqrt{2}}(|\alpha\beta\rangle + |\beta\alpha\rangle))
=
(|20\rangle,|02\rangle,|\alpha\beta\rangle,|\beta\alpha\rangle)
\begin{bmatrix}
1 & 0 & 0 & 0 \\
0 & 1 & 0 & 0 \\
0 & 0 & \frac{1}{\sqrt{2}} & \frac{1}{\sqrt{2}} \\
0 & 0 & -\frac{1}{\sqrt{2}} &  \frac{1}{\sqrt{2}}
\end{bmatrix}.
\end{eqnarray}
Then, the relation between the transformed spin eigenfunctions
with the original ones can be found as
\begin{eqnarray}
&&(|\tilde{2}0\rangle,|0\tilde{2}\rangle,\frac{1}{\sqrt{2}}(|\tilde{\alpha}\tilde{\beta}\rangle
-|\tilde{\beta}\tilde{\alpha}\rangle),
\frac{1}{\sqrt{2}}(|\tilde{\alpha}\tilde{\beta}\rangle + |\tilde{\beta}\tilde{\alpha}\rangle)) \nonumber\\
&=&
(|20\rangle,|02\rangle,\frac{1}{\sqrt{2}}(|\alpha\beta\rangle-|\beta\alpha\rangle),
\frac{1}{\sqrt{2}}(|\alpha\beta\rangle + |\beta\alpha\rangle)) 
\begin{bmatrix}
\cos^2\theta & \sin^2\theta & -\sqrt{2}\cos\theta\sin\theta & 0 \\
\sin^2\theta & \cos^2\theta & \sqrt{2}\cos\theta\sin\theta & 0 \\
\sqrt{2}\cos\theta\sin\theta & -\sqrt{2}\cos\theta\sin\theta & \cos(2\theta) & 0 \\
0 & 0 & 0 & 1 \\
\end{bmatrix}.\label{eq:rotateH11}
\end{eqnarray}
To apply Eq. \eqref{eq:rotateH11} in computing the rotated MPS with the SU(2) symmetry, the recoupling of angular momenta shown in Fig. \ref{fig:recoupling} is needed additionally. 
Specifically, the two-dot wavefunction in the spin coupling mode on the left can be
first transformed into the spin coupling mode on the right, then the wavefunction is
rotated as in Eq. \eqref{eq:WavefunctionRotation} using spin-adapted transformations like Eq. \eqref{eq:rotateH11}, and finally the rotated wavefunction is transformed
back to the original spin coupling for subsequent calculations.

\subsection{Other implementation details}
\begin{itemize}
\item Line 3: local optimization of $\theta$. To minimize $S_{1/2}[k]$
with the rotated wavefunction, we use the BOBYQA algorithm by Powell\cite{BOBYQA}
implemented in the \textsc{NLopt} package\cite{NLopt},
which performs derivative-free bound-constrained optimization using an iteratively constructed quadratic approximation for the objective function. The parameter $\theta$
is constrained to the interval $[0,\pi]$.

\item Line 4: integral transformation. Given $\mathbf{U}$, the molecular integrals
are updated by
\begin{eqnarray}
\tilde{h}_{pq} &=& \sum_{p'q'} h_{p'q'}U_{p'p}^* U_{q'q}, \\
\widetilde{\langle pq|rs\rangle}
&=&
\sum_{p'q'r's'} \langle p'q'|r's'\rangle U_{p'p}^*U_{q'q}^* U_{r'r}U_{s's}.
\end{eqnarray}

\item Line 8: the choice of $\varepsilon$. The threshold $\epsilon$ is a hyperparameter
that can be set to a larger value to enable more aggressive moves, which may increase the energy but lower the entropy. This can lead to a final MPS with lower $S_{\mathrm{tot}}$, but with higher energy than that obtained with a smaller value of $\varepsilon$ used here.
\end{itemize}

\section{Detailed results for iron-sulfur clusters}
\subsection{Computational details}
\begin{figure}[H]
\centering
\includegraphics[width=0.8\textwidth]{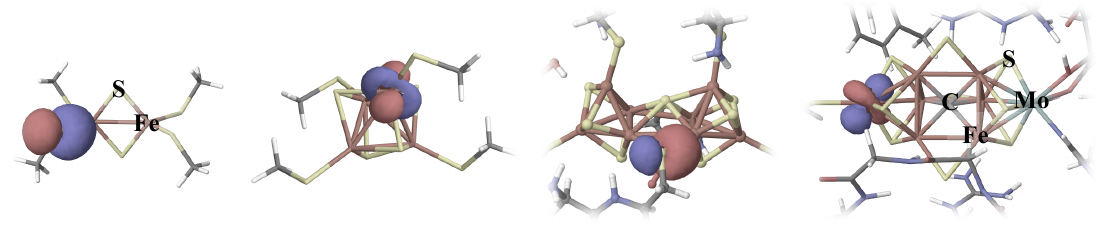}
\caption{Structures of [2Fe-2S], [4Fe-4S], P-cluster, and FeMoco, with an initial localized active orbital illustrated for each cluster.
}\label{fig:structures}
\end{figure}

We employed the previously constructed active space models 
for iron-sulfur clusters\cite{li2017spin,li2019electronic,li2019electronic2}
composed of LMOs, which include Fe $3d$, S $3p$, and bonding ligand orbitals as well as Mo $4d$ and the central C $2p$ specifically for FeMoco\cite{li2017spin,li2019electronic,li2019electronic2} .
Geometric structures and LMOs are illustrated in Fig. \ref{fig:structures} for [2Fe-2S], [4Fe-4S], P-cluster, and FeMoco.

The DMRG calculations were performed using the following Hamiltonian defined for the active orbitals,
\begin{eqnarray}
\hat{H}_{\mathrm{act}}&=&
\sum_{xy}\sum_{\sigma\in\{\alpha,\beta\}}
f_{xy}a_{x\sigma}^\dagger a_{y\sigma}
+\frac{1}{2}
\sum_{wxyz}\sum_{\sigma,\tau\in\{\alpha,\beta\}}
\langle wx|yz\rangle
a_{w\sigma}^\dagger a_{x\tau}^\dagger
a_{z\tau}a_{y\sigma}+E_{\mathrm{core}},\label{eq:CASCI}
\end{eqnarray}
where the effective one-electron integrals are defined by
$f_{xy} = h_{xy}+\sum_{i}(2\langle xi|yi\rangle-\langle xi|iy\rangle)$.
The indices $w,x,y,z$ label active orbitals, and the indices $i,j$ label doubly occupied
core orbitals. The core energy $E_{\mathrm{core}}$ is defined as the total energy
of the doubly occupied core orbitals plus the nuclear repulsion energy $E_{\mathrm{nuc}}$,
viz., $E_{\mathrm{core}} = 2\sum_{i} h_{ii}+\sum_{ij}(2\langle ij|ij\rangle-\langle ij|ji\rangle)+E_{\mathrm{nuc}}$. The corresponding molecular integrals $f_{xy}$ and $\langle wx|yz\rangle$ are available on Github\cite{linkToFCIDUMPfe2fe4,linkToFCIDUMPpclusters,linkToFCIDUMPfemoco}.

For the [2Fe-2S], [4Fe-4S], and P-cluster, the singlet ($S=0$) ground state is targeted,
while for the FeMoco, the quartet ($S=3/2$) ground state is 
studied, using spin-adapted DMRG implemented in the \textsc{FOCUS} package\cite{li2021expressibility,xiang2024distributed}. 
While the randomized orbital optimization and most of the subsequent DMRG calculations were carried out using a single CPU node, large-scale DMRG calculations for P-cluster and FeMoco were performed using four GPU nodes, each equipped with eight NVIDIA A100 40GB GPUs.
The DMRG results obtained using the initial LMO basis and the EMO basis are summarized
in Tables \ref{tab:confs}-\ref{tab:femoco}. In each table,
$E_{\mathrm{core}}$ is presented in the caption, the energy without the core
energy denoted by $E$ is tabulated for different $D$, and
the corresponding total energy can be computed by $E_{\mathrm{tot}}=\langle\Psi|\hat{H}_{\mathrm{act}}|\Psi\rangle = E + E_{\mathrm{core}}$ in order to compare with the energies reported in previous
studies\cite{li2017spin,li2019electronic,lee2023evaluating,ollitrault2024enhancing}.

\subsection{Additional results for randomized global orbital optimization}

\begin{figure}[H]
\centering
\includegraphics[width=0.8\textwidth]{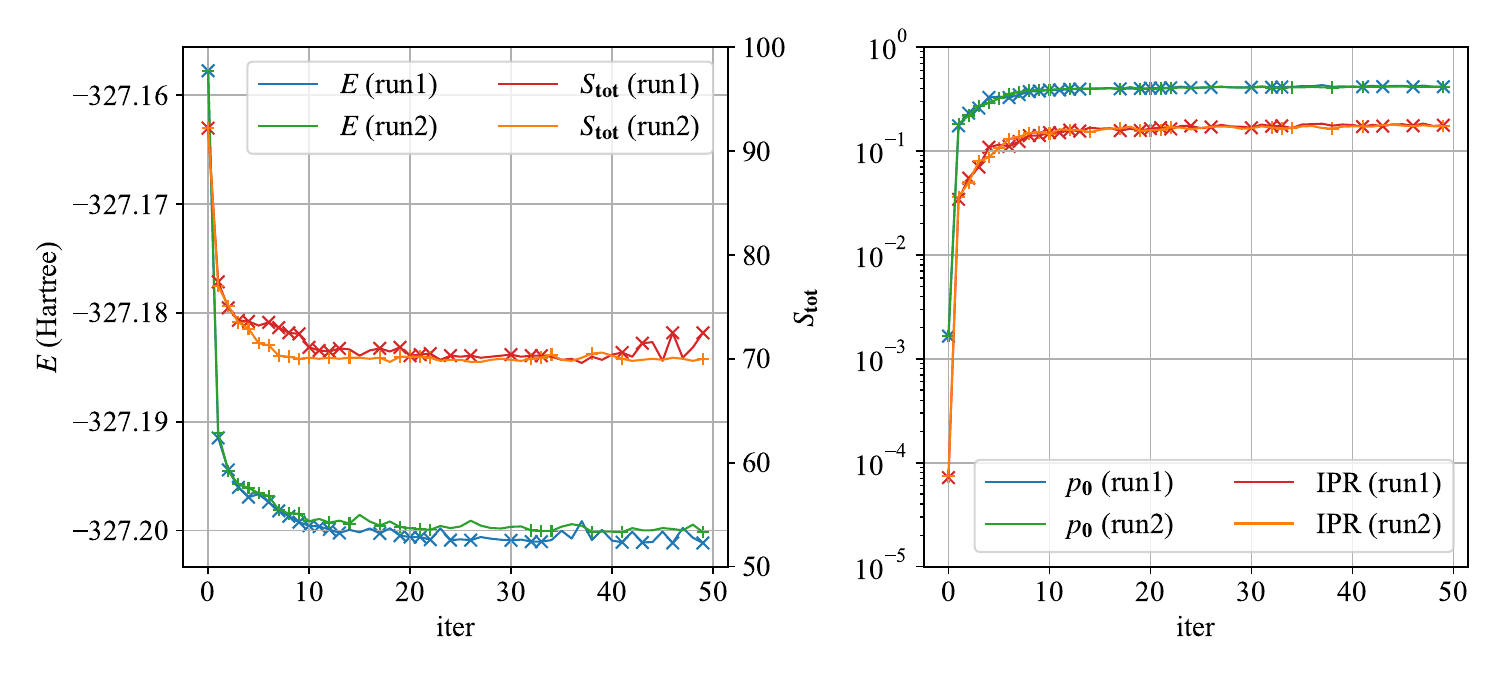}
\caption{
Randomized orbital optimization process for 
the (54e,36o) active space model of the [4Fe-4S]
cluster using $D=100$ and $N_{\mathrm{max}}=50$.
Upper: $E$ (in Hartree) and $S_{\mathrm{tot}}$. Lower: the largest weight of CSF ($p_0^\CSF$) and the inverse participation ratio (IPR).
It is seen that the optimized values for $p_{0}$ obtained from two independent runs using different seeds are in good agreement. The cross and plus markers indicate the accepted steps in the first and second runs, respectively.
}\label{fig:fe4s4}
\end{figure}


\subsection{DMRG results using LMO and EMO}


\begin{table}[H]
\caption{Leading configurations for iron-sulfur clusters found at the largest
bond dimension used in this work. The leftmost character corresponds to the first orbital.
For both CSF and DET, \texttt{2} and \texttt{0} represent
double and zero occupancies, respectively.
For CSF, \texttt{u} and \texttt{d} represent the spin coupling with the total spin plus 1/2
and -1/2, respectively.
For DET, \texttt{a} and \texttt{b} represent $\alpha$ and
$\beta$ electrons, respectively,
}\label{tab:confs}
\begin{tabular}{cccc}
\hline\hline
molecule & $D$ & basis & leading configuration \\
\hline
[2Fe-2S] & 3000 & LMO CSF &
\texttt{22uuuuu222222ddddd22} \\
&& LMO DET &
\texttt{22aaaaa222222bbbbb22} \\
&& EMO CSF &
\texttt{22uu2u2u2udd222ddd22} \\
&& EMO DET &
\texttt{22aa2a2a2abb222bbb22} \\
\protect[4Fe-4S] & 8000 & LMO CSF &
\texttt{22uuu2uuuuuu222222222222dddddd2ddd22} \\
&& LMO DET &
\texttt{22aaa2aaaaaa222222222222bbbbbb2bbb22} \\
&& EMO CSF &
\texttt{2222uuu22u2uu2u2uu2d2ddddd222d2dd222} \\
&& EMO DET &
\texttt{2222aaa22a2aa2a2aa2b2bbbbb222b2bb222} \\
P-cluster & 8000 & LMO CSF &
\texttt{2222uuu2udddd2222222222u2uuuuuu2u2222222ddd2dd2ddd222u2uuu2222222ddd2d222} \\
&& LMO DET &
\texttt{2222aaa2abbbb2222222222b2bbbbbb2b2222222aaa2aa2aaa222a2aaa2222222bbb2b222} \\
&& EMO CSF &
\texttt{22222222uu2u2udd222dduuuu222uuuu22222222dd22d2du2ududud22d2dd2dd222222222} \\
&& EMO DET &
\texttt{22222222aa2a2abb222bbaaaa222aaaa22222222bb22b2ba2ababab22b2bb2bb222222222} \\
FeMoco & 10000
& LMO CSF &
\texttt{22200uu2u22d222d2duuu22dddu2duuu2222222222222ddduu2uuuuddddd2222222222duuu22} \\
&& LMO DET &
\texttt{22200bb2022222222aaaa22bbbbaa2aa222222a222222bbbbb2bbbbaaaa2222222222aaaaa22} \\
&& EMO CSF &
\texttt{222202022220202uduuu2u222222uudd2d2du22d22222dduu2uuuud22d2d222d222dduuu2222} \\
&& EMO DET & \texttt{222202022220202baaaa2a222222aabb2b2ba22b22222bbbb2bbbba22a2a222a222aaaaa2222} \\
\hline\hline
\end{tabular}
\end{table}

\begin{table}[H]
    \renewcommand{\arraystretch}{1.2}
    \setlength{\tabcolsep}{2mm}
    \centering

\caption{DMRG results for the (30e,20o) active space model of the [2Fe-2S] cluster using LMOs and EMOs optimized with $N_{\mathrm{max}}=50$ using $D=100$.
The core energy is $E_{\mathrm{core}}=-4976.265324$ Hartree. $c_0^{\CSF}$/$c_0^{\DET}$: 
the largest absolute value of the CSF/DET coefficients.
$p_0^{CSF}$/$p_0^{\DET}$: the square of  $c_0^{\CSF}$/$c_0^{\DET}$.
Values in parenthesis are the enhancement factors compared
with the corresponding $p_0$ obtained in the LMO basis.
}\label{tab:fe2s2}
\begin{tabular}{ccccccccc}
\hline\hline
basis	&	$D$	&	$E$	&	$S_{\mathrm{tot}}$	&	IPR	&	$c_0^\CSF$	&	$p_0^\CSF$				&	$c_0^\DET$	&	$p_0^\DET$				\\
\hline
LMO	&	100	&	-116.602019	&	41.7	&	3.10e-3	&	1.86e-1	&	3.47e-2				&	7.60e-2	&	5.78e-3		\\
	&	500	&	-116.605538	&	45.6	&	2.60e-3	&	1.80e-1	&	3.22e-2				&	7.33e-2	&	5.37e-3		\\
	&	1000	&	-116.605602	&	45.9	&	2.58e-3	&	1.79e-1	&	3.21e-2				&	7.32e-2	&	5.36e-3		\\
	&	2000	&	-116.605609	&	46.1	&	2.80e-3	&	1.79e-1	&	3.21e-2				&	7.32e-2	&	5.36e-3		\\
	&	3000	&	-116.605609	&	46.1	&	2.71e-3	&	1.79e-1	&	3.21e-2				&	7.32e-2	&	5.36e-3		\\
\hline
EMO($D=100$)	&	100	&	-116.603145	&	31.6	&	1.77e-1	&	6.42e-1	&	4.12e-1	(1.19e1)	&	2.62e-1	&	6.86e-2	(1.19e1)	\\
	&	500	&	-116.605528	&	35.0	&	1.63e-1	&	6.29e-1	&	3.95e-1	(1.23e1)	&	2.57e-1	&	6.58e-2	(1.23e1)	\\
	&	1000	&	-116.605599	&	35.4	&	1.61e-1	&	6.28e-1	&	3.95e-1	(1.23e1)	&	2.56e-1	&	6.57e-2	(1.23e1)	\\
	&	2000	&	-116.605608	&	35.6	&	1.62e-1	&	6.28e-1	&	3.94e-1	(1.23e1)	&	2.56e-1	&	6.57e-2	(1.23e1)	\\
	&	3000	&	-116.605609	&	35.6	&	1.63e-1	&	6.28e-1	&	3.94e-1	(1.23e1)	&	2.56e-1	&	6.57e-2	(1.23e1)	\\
%
	\hline\hline
\end{tabular}
\end{table}

\begin{table}[H]
    \renewcommand{\arraystretch}{1.2}
    \setlength{\tabcolsep}{2mm}
    \centering

\caption{
DMRG results for the (54e,36o) active space model of the [4Fe-4S] cluster using LMOs and EMOs optimized with $N_{\mathrm{max}}=50$. The core energy is $E_{\mathrm{core}}=-8105.560390$ Hartree.
For other explanations, see Table \ref{tab:fe2s2}.
}\label{tab:fe4s4}
\begin{tabular}{ccccccccc}
\hline\hline
basis	&	$D$	&	$E$	&	$S_{\mathrm{tot}}$	&	IPR	&	$c_0^\CSF$	&	$p_0^\CSF$				&	$c_0^\DET$	&	$p_0^\DET$				\\
\hline
LMO	&	100	&	-327.157422	&	91.1	&	7.75e-5	&	4.19e-2	&	1.75e-3				&	1.32e-2	&	1.75e-4		\\
	&	500	&	-327.214060	&	117.2	&	2.45e-5	&	2.67e-2	&	7.11e-4				&	8.43e-3	&	7.11e-5		\\
	&	1000	&	-327.226723	&	128.0	&	1.67e-5	&	2.32e-2	&	5.38e-4				&	7.33e-3	&	5.38e-5		\\
	&	2000	&	-327.234789	&	137.9	&	1.25e-5	&	2.06e-2	&	4.24e-4				&	6.51e-3	&	4.24e-5		\\
	&	3000	&	-327.237975	&	143.3	&	1.14e-5	&	1.94e-2	&	3.77e-4				&	6.14e-3	&	3.77e-5		\\
	&	4000	&	-327.239683	&	146.8	&	1.02e-5	&	1.88e-2	&	3.52e-4				&	5.93e-3	&	3.51e-5		\\
	&	5000	&	-327.240749	&	149.3	&	9.41e-6	&	1.83e-2	&	3.36e-4				&	5.80e-3	&	3.36e-5		\\
	&	6000	&	-327.241478	&	151.3	&	9.35e-6	&	1.80e-2	&	3.25e-4				&	5.71e-3	&	3.26e-5		\\
&	7000	&	-327.242005	&	153.0	&	9.53e-6	&	1.79e-2	&	3.21e-4				&	5.67e-3	&3.21e-5				\\
&	8000	&	-327.242405	&	154.3	&	9.07e-6	&	1.78e-2	&	3.17e-4				&	5.63e-3	&3.17e-5				\\	
\hline
EMO($D=100$)	&	100	&	-327.200321	&	73.0	&	1.78e-1	&	6.45e-1	&	4.16e-1	(2.37e2)	&	2.04e-1	&	4.16e-2	(2.37e2)	\\
	&	500	&	-327.227047	&	96.8	&	9.61e-2	&	5.53e-1	&	3.06e-1	(4.30e2)	&	1.75e-1	&	3.06e-2	(4.30e2)	\\
	&	1000	&	-327.234198	&	107.3	&	6.96e-2	&	5.15e-1	&	2.65e-1	(4.93e2)	&	1.63e-1	&	2.65e-2	(4.93e2)	\\
	&	2000	&	-327.238697	&	116.1	&	5.74e-2	&	4.85e-1	&	2.35e-1	(5.54e2)	&	1.53e-1	&	2.35e-2	(5.54e2)	\\
	&	3000	&	-327.240564	&	120.9	&	4.94e-2	&	4.71e-1	&	2.21e-1	(5.87e2)	&	1.49e-1	&	2.21e-2	(5.87e2)	\\
	&	4000	&	-327.241589	&	124.0	&	4.37e-2	&	4.62e-1	&	2.14e-1	(6.08e2)	&	1.46e-1	&	2.14e-2	(6.08e2)	\\
	&	5000	&	-327.242233	&	126.3	&	4.47e-2	&	4.56e-1	&	2.08e-1	(6.20e2)	&	1.44e-1	&	2.08e-2	(6.20e2)	\\
	&	6000	&	-327.242679	&	128.1	&	4.31e-2	&	4.52e-1	&	2.04e-1	(6.28e2)	&	1.43e-1	&	2.04e-2	(6.26e2)	\\
&	7000	&	-327.243007	&	129.7	&	3.81e-2	&	4.49e-1	&	2.01e-1	(6.27e2)	&	1.42e-1	&2.01e-2	(6.27e2)	\\
&	8000	&	-327.243261	&	131.0	&	3.99e-2	&	4.46e-1	&	1.98e-1	(6.26e2)	&	1.41e-1	&1.99e-2	(6.26e2)	\\	
%
	\hline\hline
\end{tabular}
\end{table}

\begin{table}[H]
    \renewcommand{\arraystretch}{1.2}
    \setlength{\tabcolsep}{2mm}
    \centering

\caption{
DMRG results for the (114e,73o) active space model of the P-cluster using LMOs and EMOs optimized with $N_{\mathrm{max}}=100$. The core energy is $E_{\mathrm{core}}=-16416.706473$ Hartree.
For other explanations, see Table \ref{tab:fe2s2}.
}\label{tab:pcluster}
\begin{tabular}{ccccccccc}
\hline\hline
basis	&	$D$	&	$E$	&	$S_{\mathrm{tot}}$	&	IPR	&	$c_0^\CSF$	&	$p_0^\CSF$				&	$c_0^\DET$	&	$p_0^\DET$				\\
\hline
LMO	&	100	&	-1075.372662	&	171.0	&	5.32e-6	&	3.45e-2	&	1.19e-3				&	3.57e-3	&	1.28e-5		\\
	&	500	&	-1075.459722	&	226.5	&	3.61e-6	&	3.67e-2	&	1.35e-3				&	7.31e-3	&	5.34e-5		\\
	&	1000	&	-1075.483101	&	253.4	&	2.12e-6	&	3.06e-2	&	9.34e-4				&	1.10e-2	&	1.22e-4		\\
	&	2000	&	-1075.499669	&	278.5	&	4.86e-7	&	2.21e-2	&	4.88e-4				&	1.12e-2	&	1.25e-4		\\
	&	3000	&	-1075.507308	&	292.9	&	2.25e-7	&	1.66e-2	&	2.76e-4				&	9.59e-3	&	9.19e-5		\\
	&	4000	&	-1075.511586	&	302.2	&	1.25e-7	&	1.38e-2	&	1.90e-4				&	8.53e-3	&	7.28e-5		\\
	&	5000	&	-1075.514333	&	308.5	&	1.82e-7	&	1.24e-2	&	1.54e-4				&	8.00e-3	&	6.40e-5		\\
	&	6000	&	-1075.516286	&	313.4	&	1.24e-7	&	1.16e-2	&	1.35e-4				&	7.72e-3	&	5.96e-5		\\
	
&	7000	&	-1075.517731	&	317.4	&	3.69e-8	&	1.11e-2	&	1.23e-4				&	7.54e-3	&	5.69e-5				\\
&	8000	&	-1075.518856	&	320.8	&	9.61e-8	&	1.07e-2	&	1.14e-4				&	7.41e-3	&	5.49e-5				\\

\hline
EMO($D=100$)	&	100	&	-1075.452082	&	125.7	&	8.63e-2	&	5.39e-1	&	2.90e-1	(2.44e2)	&	1.28e-1	&	1.63e-2	(1.27e3)	\\
	&	500	&	-1075.488247	&	175.8	&	2.16e-2	&	3.69e-1	&	1.36e-1	(1.01e2)	&	1.18e-1	&	1.40e-2	(2.62e2)	\\
	&	1000	&	-1075.498670	&	201.1	&	7.03e-3	&	2.83e-1	&	8.03e-2	(8.59e1)	&	1.02e-1	&	1.05e-2	(8.62e1)	\\
	&	2000	&	-1075.507117	&	229.0	&	2.16e-3	&	2.04e-1	&	4.16e-2	(8.51e1)	&	7.93e-2	&	6.29e-3	(5.03e1)	\\
	&	3000	&	-1075.511554	&	245.7	&	6.78e-4	&	1.58e-1	&	2.49e-2	(9.04e1)	&	6.35e-2	&	4.03e-3	(4.38e1)	\\
	&	4000	&	-1075.514302	&	255.9	&	3.84e-4	&	1.34e-1	&	1.79e-2	(9.40e1)	&	5.49e-2	&	3.02e-3	(4.14e1)	\\
	&	5000	&	-1075.516178	&	262.8	&	2.92e-4	&	1.20e-1	&	1.45e-2	(9.41e1)	&	5.01e-2	&	2.51e-3	(3.92e1)	\\
	&	6000	&	-1075.517544	&	268.0	&	2.05e-4	&	1.12e-1	&	1.25e-2	(9.29e1)	&	4.71e-2	&	2.22e-3	(3.72e1)	\\
	
&	7000	&	-1075.518598	&	272.2	&	1.66e-4	&	1.06e-1	&	1.13e-2	(9.17e1)	&	4.50e-2	&	2.02e-3	(3.56e1)	\\
&	8000	&	-1075.519439	&	275.8	&	1.58e-4	&	1.02e-1	&	1.04e-2	(9.05e1)	&	4.34e-2	&	1.88e-3	(3.43e1)	\\
	\hline\hline
\end{tabular}
\end{table}

\begin{table}[H]
    \renewcommand{\arraystretch}{1.2}
    \setlength{\tabcolsep}{2mm}
    \centering

\caption{
DMRG results for the (113e,76o) active space model of the FeMoco using LMOs
and EMOs optimized with $N_{\mathrm{max}}=150$. The core energy is $E_{\mathrm{core}}
=-21021.214012$ Hartree.
For other explanations, see Table \ref{tab:fe2s2}.
}\label{tab:femoco}
\begin{tabular}{ccccccccc}
\hline\hline
basis	&	$D$	&	$E$	&	$S_{\mathrm{tot}}$	&	IPR	&	$c_0^\CSF$	&	$p_0^\CSF$				&	$c_0^\DET$	&	$p_0^\DET$				\\
\hline
LMO	&	100	&	-1118.482507	&	217.9	&	4.33e-10	&	1.53e-3	&	2.34e-6				&	1.10e-3	&	1.21e-6				\\
	&	500	&	-1118.820009	&	290.8	&	6.05e-12	&	4.25e-4	&	1.80e-7				&	4.18e-4	&	1.74e-7				\\
	&	1000	&	-1118.920234	&	321.9	&	1.66e-12	&	2.93e-4	&	8.57e-8				&	3.57e-4	&	1.27e-7				\\
	&	2000	&	-1118.996538	&	351.8	&	1.66e-12	&	1.99e-4	&	3.98e-8				&	3.05e-4	&	9.28e-8				\\
	&	3000	&	-1119.033406	&	368.7	&	2.34e-13	&	1.70e-4	&	2.88e-8				&	3.03e-4	&	9.17e-8				\\
	&	4000	&	-1119.056419	&	380.2	&	2.66e-13	&	1.51e-4	&	2.29e-8				&	2.88e-4	&	8.29e-8				\\
	&	5000	&	-1119.072124	&	388.7	&	2.03e-13	&	1.38e-4	&	1.90e-8				&	2.74e-4	&	7.48e-8				\\
	&	6000	&	-1119.083773	&	395.5	&	1.24e-13	&	1.28e-4	&	1.63e-8				&	2.61e-4	&	6.81e-8				\\
	&	7000	&	-1119.092880	&	401.0	&	1.38e-13	&	1.19e-4	&	1.43e-8				&	2.50e-4	&	6.27e-8				\\
	&	8000	&	-1119.100234	&	405.7	&	6.68e-14	&	1.15e-4	&	1.32e-8				&	2.42e-4	&	5.83e-8				\\
	
&	9000	&	-1119.106330	&	409.8	&	1.91e-13	&	1.11e-4	&	1.23e-8				&	2.35e-4	&	5.50e-8				\\
&	10000	&	-1119.111498	&	413.4	&	1.67e-13	&	1.07e-4	&	1.15e-8				&	2.28e-4	&	5.22e-8				\\
	
\hline
EMO($D=100$)	&	100	&	-1118.868657	&	173.9	&	1.04e-3	&	1.62e-1	&	2.61e-2	(1.12e4)	&	1.00e-1	&	1.00e-2	(8.28e3)	\\
	&	500	&	-1119.005819	&	240.4	&	1.25e-4	&	9.34e-2	&	8.73e-3	(4.84e4)	&	7.07e-2	&	5.00e-3	(2.87e4)	\\
	&	1000	&	-1119.050973	&	269.3	&	5.58e-5	&	7.23e-2	&	5.22e-3	(6.10e4)	&	5.98e-2	&	3.57e-3	(2.81e4)	\\
	&	2000	&	-1119.087453	&	297.3	&	2.20e-5	&	5.64e-2	&	3.18e-3	(8.00e4)	&	4.99e-2	&	2.49e-3	(2.69e4)	\\
	&	3000	&	-1119.105478	&	313.3	&	1.05e-5	&	4.91e-2	&	2.41e-3	(8.35e4)	&	4.50e-2	&	2.03e-3	(2.21e4)	\\
	&	4000	&	-1119.116896	&	324.4	&	8.57e-6	&	4.48e-2	&	2.01e-3	(8.77e4)	&	4.20e-2	&	1.77e-3	(2.13e4)	\\
	&	5000	&	-1119.124937	&	332.8	&	5.08e-6	&	4.20e-2	&	1.76e-3	(9.27e4)	&	3.99e-2	&	1.59e-3	(2.13e4)	\\
	&	6000	&	-1119.131007	&	339.5	&	3.72e-6	&	3.98e-2	&	1.59e-3	(9.75e4)	&	3.83e-2	&	1.46e-3	(2.15e4)	\\
	&	7000	&	-1119.135803	&	345.1	&	4.81e-6	&	3.81e-2	&	1.45e-3	(1.02e5)	&	3.70e-2	&	1.37e-3	(2.18e4)	\\
	&	8000	&	-1119.139701	&	349.9	&	2.47e-6	&	3.68e-2	&	1.35e-3	(1.03e5)	&	3.59e-2	&	1.29e-3	(2.21e4)	\\
	
&	9000	&	-1119.142944	&	354.0	&	3.39e-6	&	3.57e-2	&	1.27e-3	(1.04e5)	&	3.50e-2	&	1.23e-3	(2.23e4)	\\
&	10000	&	-1119.145684	&	357.6	&	2.72e-6	&	3.48e-2	&	1.21e-3	(1.05e5)	&	3.43e-2	&	1.17e-3	(2.25e4)	\\

	\hline\hline
\end{tabular}
\end{table}

\subsection{Comparison with $p_0$ from previous works}
We document the values of $p_0$ from Refs. \cite{ollitrault2024enhancing,lee2023evaluating} 
used for comparison in Fig. \ref{fig:fescluster}.
Note that the active spaces for the [2Fe-2S] and [4Fe-4S] clusters in Refs. \cite{ollitrault2024enhancing,lee2023evaluating} were regenerated by the authors following the procedure in Ref. \cite{li2017spin}. Therefore, there can be small difference in the results. Another source of the difference can come from the fact that
DMRG calculations are usually carried out with small noises, which can also lead
to slightly different $p_0$.

For the [2Fe-2S] cluster, Ref. \cite{ollitrault2024enhancing} reports 
$p_0^\CSF$(LMO)=3.21e-2, $p_0^\DET$(LMO)=5.35e-3, and
$p_0^\DET$(OPT)=4.53e-2 for the optimized (OPT)
orbitals using their algorithm. The LMO results agree well with our results in Table
\ref{tab:fe2s2}. Our result for $p_0^\DET$=6.57e-2 using EMO($D=100$) is larger than
$p_0^\DET$(OPT) by a factor of 1.4.

For the [4Fe-4S] cluster, Ref. \cite{ollitrault2024enhancing} reports
$p_0^\CSF$(LMO)=2.94e-4,
$p_0^\DET$(LMO)=2.94e-5, and
$p_0^\DET$(OPT)=2.67e-3 at $D=8000$.
The corresponding data at $D=7000$ in Ref. \cite{lee2023evaluating} 
are $p_0^\CSF$(LMO)=2.34e-4 and $p_0^\DET$(LMO)=2.34e-5.
Our results at $D=8000$ shown in Table \ref{tab:fe4s4}
are $p_0^\CSF$(LMO)=3.17e-4 and
$p_0^\DET$(LMO)=3.17e-5, which agree reasonably well with the reported results.
Our result $p_0^\DET$=1.99e-2 using EMO($D=100$) is about 7.5 times
larger than $p_0^\DET$(OPT).

For the P-cluster, the results obtained with the largest $D$ in Ref. \cite{lee2023evaluating} are $p_0^\CSF$(LMO)=1.51e-4 and $p_0^\DET$(LMO)=6.32e-5 at $D=5000$, which are very close to our results in Table \ref{tab:pcluster}
obtained at $D=5000$.

For the FeMoco, the results obtained with the largest $D$ in Ref. \cite{lee2023evaluating} are $p_0^\CSF$(LMO)=1.35e-7 at $D=6000$ and
$p_0^\DET$(LMO)=9.76e-9 at $D=7000$.
Our results in Table \ref{tab:femoco} are about 10 times smaller for 
$p_0^\CSF$(LMO) but 10 times larger for $p_0^\DET$(LMO).
Such large difference is more likely due to the convergence
to different low-lying competing states as discussed in the main text.

\clearpage
\section{Additional results for 2D Hubbard model}
\begin{figure}[H]
\centering
\includegraphics[width=0.6\textwidth]{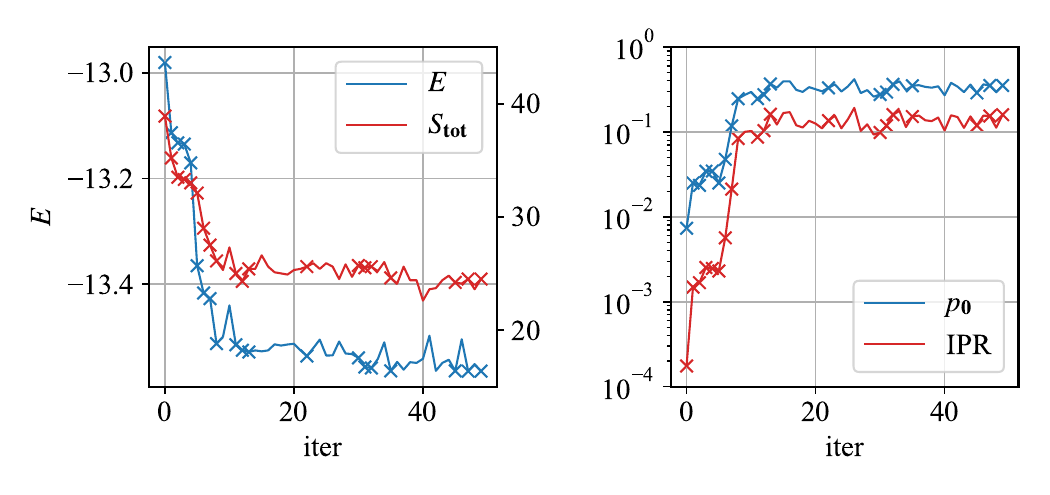}
\caption{
Randomized orbital optimization process with $D=100$ and $N_{\mathrm{max}}=50$
for the 4-by-4 Hubbard model ($U=4$). 
Left: $E$ (in Hartree) and $S_{\mathrm{tot}}$ (inset); Right: $p_0^\CSF$ and IPR.
The cross markers indicate the accepted steps.
}\label{fig:hubbard2d-opt}
\end{figure}

Apart from iron-sulfur clusters, we also benchmark our algorithm on the two-dimensional Hubbard model\cite{leblanc2015solutions,arovas2022hubbard,qin2022hubbard},
with the Hamiltonian given by
\begin{eqnarray}
\hat{H} = \sum_{\langle i,j\rangle}\sum_{\sigma}
-t (\hat{a}_{i\sigma}^\dagger \hat{a}_{j\sigma}
+ \hat{a}_{j\sigma}^\dagger \hat{a}_{i\sigma})
+
U \sum_{i} \hat{n}_{i\alpha}\hat{n}_{i\beta},\label{eq:Hubbard}
\end{eqnarray}
where $\hat{n}_{i\sigma}=
\hat{a}^\dagger_{i\sigma}\hat{a}_{i\sigma}$,
and $\langle i,j\rangle$ denotes the summation over nearest-neighbor sites.
We consider the 4-by-4 system under the periodic boundary condition (PBC)
with $t=1$ and $U=4$.
The initial orbital basis is the site basis with the 
zigzag ordering for the DMRG calculations. 
Figure \ref{fig:hubbard2d-opt} summarizes the results obtained using the randomized orbital optimization with $D=100$ and $N_{\max}=50$ iterations.
We observe a rapid decrease in both $E$ and $S_{\mathrm{tot}}$ during 
the first 10 iterations. Correspondingly, $p_0^\CSF$ increases dramatically from 0.0073 to 0.35, by about a factor of 48, while the IPR also increases from $\mathcal{O}(10^{-4})$ to $\mathcal{O}(10^{-1})$. 
Table \ref{tab:hubbard2d} demonstrates that, as $D$ increases, 
$p_0$ for CSF/DET in the EMO basis remain almost unchanged, 
while those in the site basis decrease.
This results in a larger enhancement of $p_0$ for larger $D$ in the EMO basis.
Furthermore, the DMRG energy in the EMO basis exhibits significantly faster convergence (by more than an order of magnitude) with increasing bond dimension $D$ than in the site basis. These findings are consistent with the findings
for iron-sulfur clusters.

\begin{table}[H]
    \renewcommand{\arraystretch}{1.2}

\centering
\caption{
DMRG results for the 4-by-4 Hubbard model ($U=4$) using site basis, NOs, and EMOs.
$\Delta E$: the energy error with respect to the exact diagonalization result
$E_{\mathrm{exact}}=-13.621855$;
$S_{\mathrm{tot}}$: total entropy;
$p_0$: the largest weight of CSF/DET.
Numbers in parenthesis are the enhancement factors
with respect to the corresponding $p_0$ in the site basis.
}\label{tab:hubbard2d}
\begin{tabular}{ccccccccccccccc}
\hline\hline
& \multicolumn{4}{c}{site basis} && 
\multicolumn{4}{c}{NO basis} && 
\multicolumn{4}{c}{EMO basis} \\
\cline{2-5}\cline{7-10}\cline{12-15}
$D$ & 
$\Delta E$& $S_{\mathrm{tot}}$ & $p_0^{\CSF}$ & $p_0^{\DET}$ &&
$\Delta E$& $S_{\mathrm{tot}}$ & $p_0^{\CSF}$ & $p_0^{\DET}$ &&
$\Delta E$& $S_{\mathrm{tot}}$ & $p_0^{\CSF}$ & $p_0^{\DET}$ \\
\hline
100	&	0.641348	&	38.9	&	7.34e-3	&	8.46e-3	&	&	1.459348	&	27.4	&	1.80e-1	(24.6)	&	1.80e-1	(21.3)	&	&	0.057459	&	24.5	&	3.53e-1	(48.1)	&	8.82e-2	(10.4)	\\
500	&	0.131875	&	48.0	&	4.97e-3	&	8.63e-3	&	&	0.442987	&	36.0	&	6.50e-2	(13.1)	&	6.50e-2	(7.5)	&	&	0.010193	&	29.0	&	3.49e-1	(70.2)	&	8.72e-2	(10.1)	\\
1000	&	0.044477	&	50.8	&	4.55e-3	&	8.34e-3	&	&	0.200157	&	39.9	&	5.22e-2	(11.5)	&	5.22e-2	(6.3)	&	&	0.002563	&	30.4	&	3.48e-1	(76.6)	&	8.71e-2	(10.4)	\\
2000	&	0.010308	&	52.5	&	4.36e-3	&	8.19e-3	&	&	0.061189	&	43.5	&	4.31e-2	(9.9)	&	4.31e-2	(5.3)	&	&	0.000444	&	31.2	&	3.48e-1	(79.8)	&	8.70e-2	(10.6)	\\
3000	&	0.003057	&	53.2	&	4.32e-3	&	8.15e-3	&	&	0.020504	&	45.4	&	4.00e-2	(9.3)	&	4.00e-2	(4.9)	&	&	0.000105	&	31.5	&	3.48e-1	(80.5)	&	8.70e-2	(10.7)	\\
4000	&	0.000800	&	53.5	&	4.31e-3	&	8.13e-3	&	&	0.006717	&	46.5	&	3.81e-2	(8.9)	&	3.81e-2	(4.7)	&	&	0.000022	&	31.6	&	3.48e-1	(80.8)	&	8.70e-2	(10.7)	\\
5000	&	0.000150	&	53.6	&	4.31e-3	&	8.13e-3	&	&	0.002130	&	47.3	&	3.73e-2	(8.7)	&	3.73e-2	(4.6)	&	&	0.000003	&	31.7	&	3.48e-1	(80.8)	&	8.70e-2	(10.7)	\\
6000	&	0.000008	&	53.6	&	4.31e-3	&	8.13e-3	&	&	0.000804	&	47.8	&	3.68e-2	(8.5)	&	3.68e-2	(4.5)	&	&	0.000000	&	31.7	&	3.48e-1	(80.8)	&	8.70e-2	(10.7)	\\
\hline
\hline
\end{tabular}
\end{table}

\begin{figure}[H]
\centering
\includegraphics[width=0.6\textwidth]{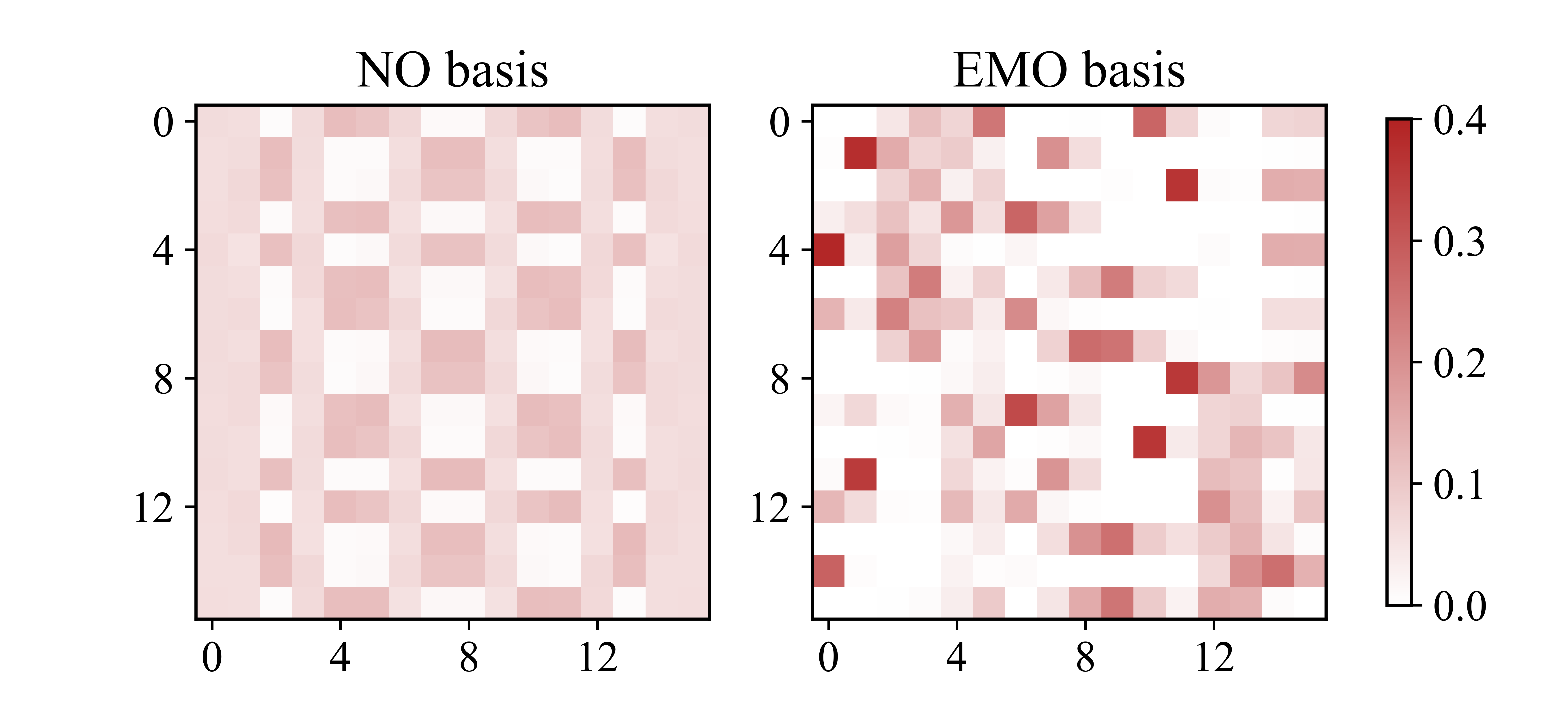}
\caption{Comparison of NOs and EMOs in terms of the square of the molecular orbital coefficients $|U_{ij}|^2$ for the 4-by-4 Hubbard model ($U=4$).}\label{fig:hubbard2d-NOvsEMO}
\end{figure}

For comparison, we also present the results obtained using natural orbitals\cite{lowdin1955quantum} (NOs) in Table \ref{tab:hubbard2d}. The NO basis is constructed using the one-body reduced density matrix (1RDM) obtained from DMRG calculations with $D=1000$ in the site basis.
We find that using NOs results in a higher $p_0$ than the site basis,
by a factor of 8.5 (4.5) at $D=6000$ for CSF (DET). 
However, the improvement in $p_0$ using EMOs over NOs is strikingly large. 
Specifically, the improvement for $p_0^\CSF$ is nearly an order of magnitude. 
Moreover, the DMRG energy using NOs is the worst among all orbital choices, 
likely due to the delocalized nature of NOs (see Fig. \ref{fig:hubbard2d-NOvsEMO}), 
making them less suited for DMRG calculations.

\end{document}